\documentclass[twocolumn]{aa}
\usepackage{graphicx,amssymb}
\newcommand \msun {\mbox{{M}$_{\odot}$}}
\newcommand \kms {\mbox{km~s$^{-1}$}}
\newcommand \degree {\mbox{$^\circ$}}

\newcommand \solar {$L_{\odot}$\ }

\begin{document}
   \title{Molecular gas and continuum emission in 3C48:\\ 
   Evidence for two merger
   nuclei?}


   \author{M.\ Krips
          \inst{1}
          \and
          A.\ Eckart\inst{1}\fnmsep
	  \and
	  R.\ Neri\inst{2}
	  \and
	  J.\ Zuther\inst{1}
	  \and
	  D.\ Downes\inst{2}
	  \and
	  J.\ Scharw\"achter\inst{1}
          }

   \offprints{M.\ Krips\\ 
              e-mail: krips@ph1.uni-koeln.de}

   \institute{I. Physikalisches Institut, Universit\"at zu K\"oln,
              Z\"ulpicher Str. 77, 50937 K\"oln
         \and
              Institut de Radio Astronomie Millim\'etrique, 
              300 rue de la Piscine, 38406 Saint Martin
              d'H\`eres, France
             }
   \date{Received ; accepted }

   \abstract{We present new interferometer observations of the
     CO(1--0) line and mm continuum emission from 3C48 --- one of the
     nearest examples of a merger activating a quasar. Our new CO data
     show that most of the CO is not in a disk around the quasar 3C48,
     but rather in a second nucleus associated with the source 3C48A
     $\sim 1''$ to the north-east, recently studied in the near-IR by
     Zuther et al.\ (2004).  This main CO source has a strong velocity
     gradient (140\,\kms\ over about 1$''$).  Our new data also show a
     second, weaker CO source at the QSO itself. At 1.2\,mm, the
     continuum emission is elongated in the direction of the radio jet
     and towards 3C48A. We model the 1.2\,mm continuum with three
     different sources in 3C48 --- the 3C48 QSO, the 3C48 jet, and the
     second nucleus 3C48A. We suggest that the unusually bright and
     extended nature of the jet may be due to its interaction with the
     second merger nucleus 3C48A.   } \maketitle
%

\begin{figure*}
\centering
\resizebox{\hsize}{!}{\rotatebox{-90}{\includegraphics{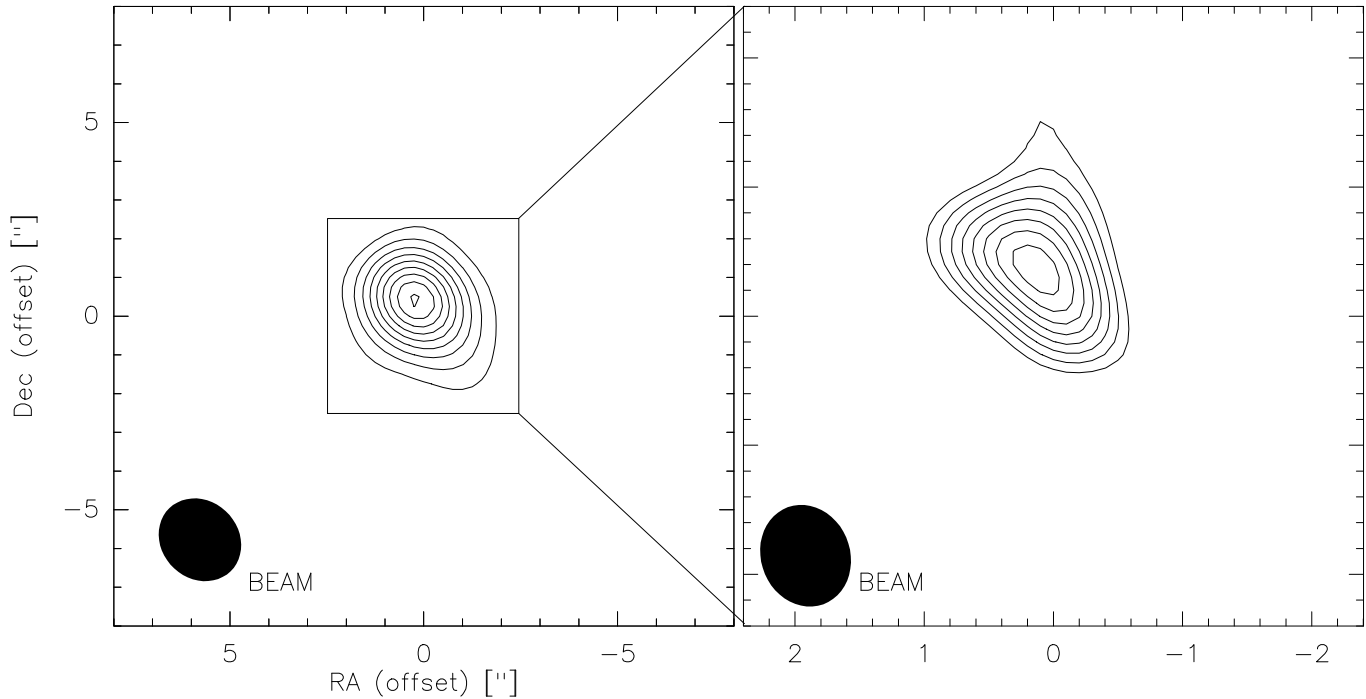}}}
\caption{Continuum maps at 3.5\,mm ({\it left}) and 1.2\,mm ({\it
right}) made with the IRAM interferometer in 2003, shown as contours.
Contour levels at 3.5\,mm are 0.025 to 0.23 by 0.025\,Jy/beam
(20$\sigma$). Contour levels at 1.2\,mm are 0.01\,Jy/beam (5$\sigma$)
to 0.044\,Jy/beam by 0.004\,Jy/beam (2$\sigma$). Both maps were made
with uniform weighting. Beams are $2.3''\times 2.0''$ at PA=44\degree
at 3.5\,mm and $0.8''\times 0.7''$ at PA=20\degree at 1.2\,mm.}
\label{contmm}
\end{figure*}

   \begin{table*}
     \centering
      \caption[]{Continuum data at 3.5\,mm and 1.2\,mm from all available
      data sets.}
     \begin{tabular}{ccccccc}
            \hline
            \hline
 Observing & RA$^{\mathrm{a}}$ & Dec.$^{\mathrm{a}}$ 
&\multicolumn{2}{c}{Continuum flux density$^{\mathrm{c}}$} 
                     &3.5\,mm beam      &Reference\\
 date	   &J2000    &J2000 &   3.5\,mm   &1.2\,mm &maj.,min.,PA\\
	   &(hh:mm:ss)&(dd:mm:ss) & (mJy) &(mJy)&($''$,\ $''$,\ $^\circ$)\\
            \hline
\multicolumn{3}{l}{{\bf Observations:}}\\
1992        &--- &---                &266 &---  
                   &$10.5''\times 8.6''$ &Scoville et al.\ (1993)\\  
1995-W97    &--- &---     &$307\pm 30$ &---  
                   &$4.3''\times 1.8''$, 34\degree &Wink et al. (1997)\\  
1995-new$^{\mathrm{b}}$ &01:37:41.29 &33:09:35.4 &$303\pm 30$ &--- 
                   &$4.0''\times 2.0''$, 38\degree &this paper\\
2003        & 01:37:41.30 &33:09:35.5 
                   &$270\pm 27$ &$82\pm 8^{\mathrm{d}}$ 
                   &$2.3''\times 2.0''$, 44\degree  &this paper\\ 
\multicolumn{3}{l}{{\bf Three-Component Model}$^{\mathrm{e}}$:}\\ 
``QSO''   & & &---   &$36\pm 8$ \\
``Jet''   & & &---   &$15\pm 4$ \\
``3C48A'' & & &---   &$25\pm 7$ \\
 \hline
\end{tabular}
 \\
\begin{list}{}{}
\item[$^{\mathrm{a}}$] Centroid of the 3.5\,mm continuum emission\\
\item[$^{\mathrm{b}}$] Our re-reduction of W97's data.\\
\item[$^{\mathrm{c}}$] Flux errors include 10\% systematic errors. \\ 
\item[$^{\mathrm{d}}$] Total
      flux of all components, beam at 1.2\,mm:
      $0.9''\times 0.7''$ at PA 15\degree. \\
\item[$^{\mathrm{e}}$] Fluxes derived by
      fitting 3 Gaussian components in the sky plane at 1.2\,mm (see
      Fig.\ref{model1mm}).
         \label{obsval}
\end{list}
   \end{table*}

\section{Introduction}

Galaxy mergers are regarded as part of the chain of events that can
lead to the activation or re-ignition of quasars.  A prime example of
this phenomenon is the radio source 3C48, one of the first
quasars to be optically identified (Matthews et al.\ 1961).  The
source is known to be surrounded by an unusually large and bright host
galaxy with a young stellar population (Kristian 1973; Boronson \& Oke
1982; 1984).  The properties of this host galaxy, the existence of a
second bright compact component, 3C48A, 1$''$ northeast of the QSO
(Stockton \& Ridgeway 1991; Zuther et al.\ 2004), the tail-like
extension to the northeast (Canalizo \& Stockton 2000), and the
richness of 3C48 in molecular gas (Scoville et al.\ 1993; Wink et al.\
1997, hereafter W97) have all been used as arguments that the activity
in 3C48 is due to a recent merger.

An uncertainty in the merger picture for 3C48 has been the unknown
nature of 3C48A. Although Zuther et al.\ (2004) recently
detected 3C48A in the near-IR, suggesting it may be a second nuclear
bulge, those authors noted that 3C48A may simply result from the
interaction of the radio jet with the surrounding interstellar medium
in the host galaxy (see also Chatzichristou et al.\ 1999).  Our new CO
results in this paper strengthen the idea that 3C48A is indeed a
second nuclear bulge, surrounded by a massive circumnuclear disk of
molecular gas.

Because two tidal tails are often seen in major mergers, the absence
of a tidal counter-tail in 3C48 has been another mystery in the merger
scenario, but Scharw\"achter et al.\ (2004) recently proposed a simple
solution for the missing tidal counter-tail.  Whereas previous papers
suggested that the tail extends from southeast to southwest (Canalizo
\& Stockton 2000), or confused a background galaxy with the
counter-tail (Boyce et al.\ 1999; Canalizo \& Stockton 2000),
Scharw\"achter et al.\ suggest that the counter-tail is in front of
the main body of the 3C48 host galaxy, running from southwest to
northeast.  They assume a similar configuration as in the
Antennae galaxies but viewed from a different angle (as in the
diagrams of Toomre \& Toomre 1972), and supported their picture with
multi-particle simulations that agree better with the observed stellar
kinematics than in previous scenarios.

The information on molecular gas in 3C48 began with 
Scoville et al.\ (1993)'s detection of CO(1--0) with the
Caltech interferometer.  Wink et al.\ (1997) then confirmed Scoville et
al.'s CO results with the IRAM Interferometer. Both groups found a
large mass of molecular gas, a few times $10^{10}$\msun.  Because the
molecular gas plays a major role in forming new stars and fueling a
black hole, further study of the molecular gas in 3C48 may help us to
understand the many steps leading to the intense nuclear activity of
quasars.

In this paper, we present new interferometer observations of the
CO(1--0) line and the 3.5 and 1.2\,mm continuum emission in 3C48.  To
increase the sensitivity, the data were combined with the earlier
measurements by Wink et al.\ (1997). The combined data set is about
twice as sensitive as the W97 data. In this paper, section~2 gives
details of the new observations. Section~3 presents the results of the
combined data sets for the mm continuum and the CO emission. In
section~4, we discuss the estimate of the molecular gas mass. In
section~5, we compare 3C48 with two other mergers, the ``Antennae''
galaxies and Arp\,220, and summarize our results in section~6.

%
\begin{figure}[!]
\centering
\resizebox{\hsize}{!}{\rotatebox{-90}{\includegraphics{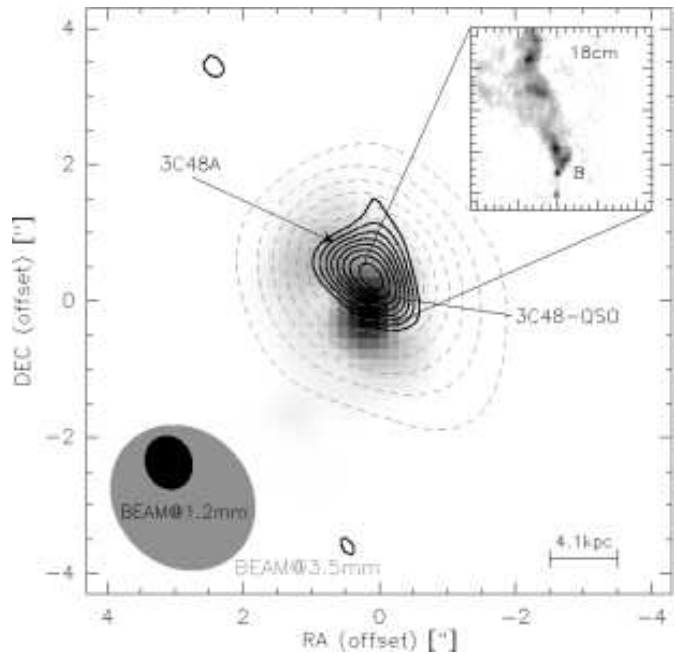}}}
\caption{Contours of the 3.5\,mm ({\it grey dashed lines}) and 1.2\,mm
  ({\it black solid lines}) continuum observed in 2003.  The contours
  are the same as in Fig.\ref{contmm}, and are superposed on the
  greyscale NIR-image (Zuther et al.\ 2004) with the QSO nucleus
  subtracted (its position agrees with the radio QSO position and is
  thus located at zero in our maps). The radio jet at 18\,cm is
  plotted in the small box ({\it upper right}; from Wilkinson et al.\
  1991; tickmarks correspond to 0.1$''$). B denotes the position of
  the QSO.}
\label{cont-ir}
\end{figure}

\section{Observations}
\label{secobs}
We used two independent data sets of the CO(1--0) emission in 3C48,
observed with the IRAM interferometer on Plateau de Bure, France.  The
first was taken by Wink et al.\ (1997) between November 1994 and
February 1995 in the interferometer's B and C configurations with 3
antennas.  We re-reduced the data from W97 (see below), and merged
them for the analysis of the CO emission with our second data
set, observed between December 2002 and March 2003 in the A and C
configurations with 6 antennas.  Following W97, we chose 3C84 and
3C111 as amplitude and phase calibrators. 3C84 was also used to
calibrate the RF bandpass. The receiver at 3.5\,mm was tuned to
84.17\,GHz, the frequency of $^{12}$CO(1--0) at $z=0.3695$, and the
receiver at 1.2\,mm to 241.39\,GHz, the frequency of $^{13}$CO(3--2)
at $z=0.3695$.  We used a total bandwidth of 580\,MHz at 3.5\,mm and
$2\times 580$\,MHz (DSB) at 1.2\,mm with a frequency resolution of
1.25\,MHz. The phase center was adopted from W97, i.e. set to 01$^{\rm
h}$37$^{\rm m}$41.3$^{\rm s}$, +33\degree09$'$35.0$''$ (J2000). This
is identical with the radio position of the QSO (e.g., Feng et al.,
2005). The total on-source integration time was
$\sim$6.5\,hours. Although the 1.2\,mm receiver was tuned to the
redshifted frequency of $^{13}$CO(3--2), no 1.2\,mm line signal was
detected to a flux limit of 3.4\,Jy\,\kms\ over the same velocity
range of $-$220 to +160\,\kms\ where the CO(1--0) is observed.  The
1.2\,mm observations were therefore used for the 3C48 continuum data.

The most critical point in merging data sets from different epochs
with each other is the flux calibration. However, for our re-reduction
of W97's data we adopted the flux calibration described in W97
resulting in an accuracy of $\sim$10\%. Except slightly different beam
sizes which can be traced back to slightly different data flagging,
our re-reduction of W97's data is consistent with the one done by
W97. For our new data, we relied on the flux monitoring which is
regularly done at the IRAM PdBI giving an accuracy of $\sim$10\% at
3.5~mm. In summary, this would yield an uncertainty of $\lesssim$20\%
of the merged data in the worst case. As will be discussed in
Section~\ref{3mmcont} \& \ref{co10}, the fluxes and positions derived
for the CO and continuum emission in the two data sets agree well with
each other within $\sim$10\% justifying thus a merging.

\begin{figure}[!]
\centering
\resizebox{\hsize}{!}{\rotatebox{-90}{\includegraphics{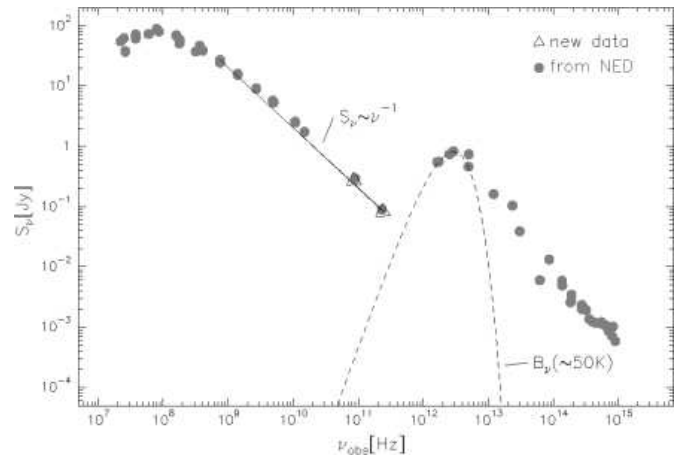}}}
\caption{Flux density spectrum of 3C48.  Grey points are from the
NASA/IPAC Extragalactic Database (NED).  The triangles at 3.5\,mm and
1.2\,mm are from our new data, and are the total flux of our continuum
fit components.  The solid line is a power-law fit to the cm-radio
synchrotron emission, and the dashed curve in the infrared
(2--10\,THz) is a model spectrum for a rest-frame dust temperature of
50\,K, a dust $\tau = 1$ at rest-frame 100\,$\mu$m, and an
optically-thin, $\nu^4$ dependence in the mm range.}
\label{sed}
\end{figure}

\begin{figure}[!]
\centering
\resizebox{\hsize}{!}{\rotatebox{-90}{\includegraphics{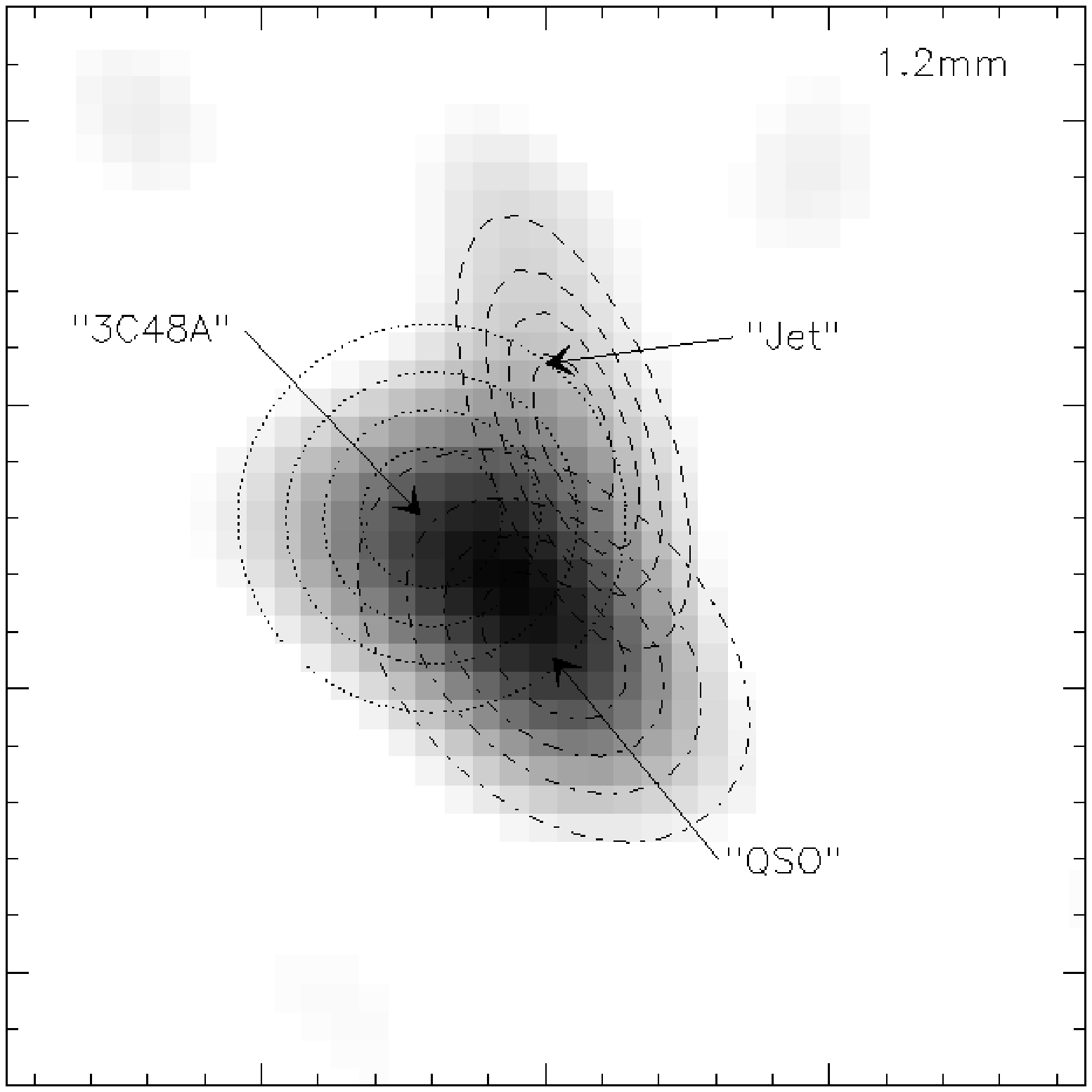}}}
\vskip -0.055cm
\resizebox{\hsize}{!}{\rotatebox{-90}{\includegraphics{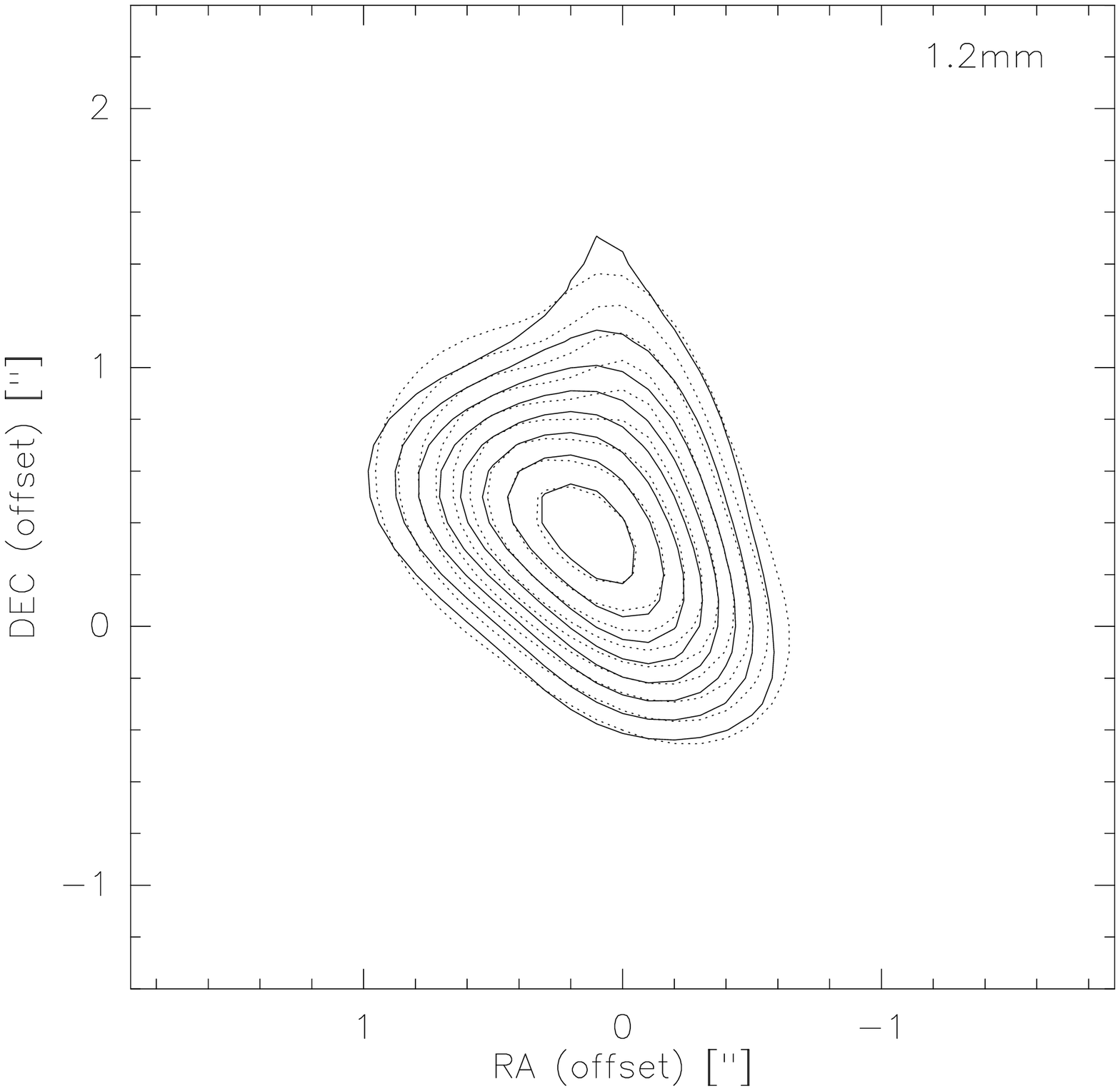}}}
\caption{ {\it Upper Panel:} Contours of a triple-Gaussian model:
``QSO'' ({\it dotted/dashed lines}), ``jet''({\it dashed lines}), and
``3C48A'' ({\it dotted lines}).  Contours are from 20 to 100\% in
steps of 20\% of the peak fluxes of each component.  {\it Lower
Panel:} The sum of the model components ({\it black dashed contours}),
compared with the observed 1.2\,mm continuum ({\it black solid
contours}); data and model contours are as in
Fig.\,\ref{cont-ir}. Data are taken from the 2003 data set only. }
\label{model1mm}
\end{figure}

\begin{figure}[!]
\centering
\resizebox{\hsize}{!}{\rotatebox{-90}{\includegraphics{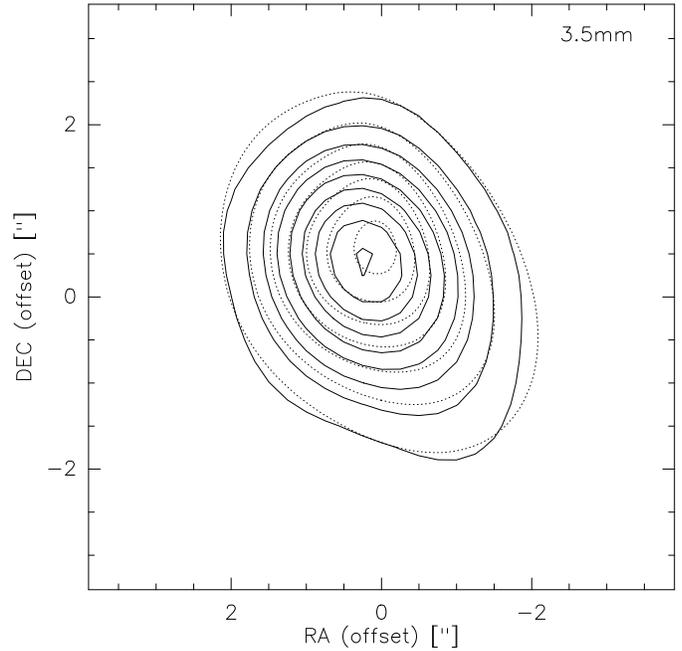}}}
\caption{Continuum emission at 3.5\,mm ({\it black solid contours};
2003 data only) overlaid with contours of a three Gaussian component
model taken from 1mm with corrected fluxes and smoothed to the 3.5\,mm
beam ({\it black dashed lines}). Contours of solid lines ({\it lower
panel}) for model and data as in Fig.\ref{cont-ir}.}
\label{model3.5mm}
\end{figure}

\section{The data}

\subsection{Continuum}

The strong mm continuum emission by itself shows 
the intense nuclear activity in 3C48. The mm continuum is 
from synchrotron radiation, not dust.

\subsubsection{The 3.5\,mm continuum}
\label{3mmcont}
We independently measured the 3.5\,mm continuum fluxes from both our
re-reduction of the 1995 data and our new 2003 data by averaging all
channels more than $\pm 380$\,\kms\ from the CO line.  The flux from
our re-reduction of the 1995 data agrees with that from W97, while the
flux from the 2003 data is 10\% lower, which is within the systematic
errors in the flux calibration, but may just be a real effect,
possibly due to intrinsic variability of 3C48 (Table~\ref{obsval}).

The position obtained from our mm-observations is 0.4$''$ north of
that given by 13 and 3.5\,cm VLBI measurements (Ma et al.\ 1998). This
difference is probably due to the VLBI observations picking out the
low-flux, milliarcsecond-scale hot spot near the quasar (see the maps
by Wilkinson et al. 1991 and by Feng et al.\ 2005), while the mm
interferometer sees mainly the high-flux, arcsecond-scale jet.  While
this position difference is about twice the astrometric uncertainty
limit of the IRAM interferometer, a comparison with the recent NIR
images of 3C48-QSO and 3C48A nevertheless suggests that the mm
continuum may be located between the two NIR features
(Fig.~\ref{cont-ir}).

The continuum emission is slightly extended in the $2''$ beam
(Fig.\ref{contmm}) at 3.5\,mm. The two nuclear components are not
resolved but the peak flux of 230\,mJy/beam is lower than the
integrated flux density of 270\,mJy, so the continuum is not a single
point source. This is also evident from the cm-radio maps and spectra
published by Feng et al.\ 2005. Most of the (cm-)radio continuum flux
comes from the radio jet and/or the northeast component 3C48A.  The
continuum flux at 3.5\,mm falls on the steep spectrum (with
$S_\nu\propto\nu^{-1}$) observed at cm wavelengths (Fig.\,\ref{sed}
and Meisenheimer et al.\ 2001).  The continuum emission at 3.5\,mm is
thus still dominated by the jet's synchrotron radiation, not thermal
dust emission.

\begin{figure*}[!]
   \centering
   \resizebox{\hsize}{!}{\rotatebox{-90}{\includegraphics{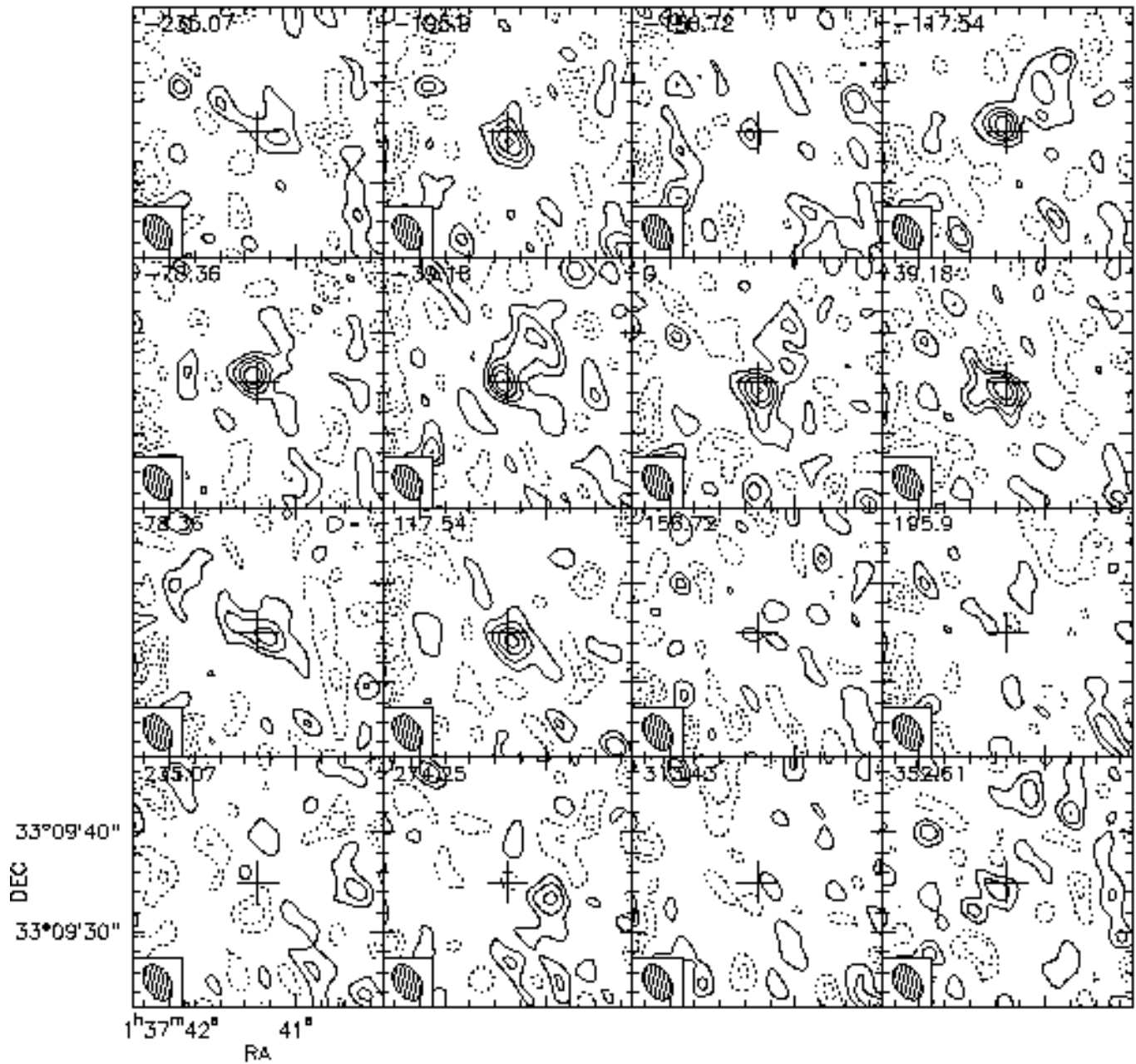}}}
   \caption{CO(1--0) channel maps from the merged 1995+2003
data. Positive and negative contours are in steps of
1.2\,mJy\,beam$^{-1}$ (1$\sigma$), up to a maximum of
6.0\,mJy\,beam$^{-1}$. The natural weighted beam is
3.6$''\times$2.5$''$ at PA=32\degree. The cross in the maps
denotes the phase center (Section~\ref{secobs}).}
   \label{co-chan}
\end{figure*}

\subsubsection{The 1.2\,mm continuum}

The 1.2\,mm continuum was observed only in the 2003 data set
(Fig.\ref{contmm}). Because we found no line emission at 1.2\,mm, we
averaged the continuum over the entire bandwidth of $2\times$580\,MHz
(DSB). The continuum was detected with a SNR of $\sim$25,
well-centered on the 3.5\,mm position.  The emission appears extended
to the northeast, toward the near-IR component 3C48A and toward the
north of the radio jet (Fig.\ref{cont-ir}).  These extensions are
similar to those on the 18\,cm MERLIN map (Akujor et al.\ 1994). This
motivated us to fit three Gaussian components (3C48A-QSO, 3C48A, and
the jet) to our 1.2\,mm map (Fig.\,\ref{model1mm}), both to explain
the unusual shape, and to estimate fluxes for each of the possible
Gaussian components to $\sim 4 \sigma$-accuracy (Table~1).
Extrapolating the cm continuum fluxes to 1.2\,mm with the steep
spectral index of $\alpha=-$1 ($S_\nu\propto \nu^\alpha$; Fig.~3)
yields a flux close to the measured flux of the QSO and jet.  That is,
even at 1.2\,mm, the optically thin synchrotron radiation from
the jet still predominates over the expected dust flux from the CO
source (by more than an order of magnitude; compare Fig.~\ref{sed}).
  
The Gaussian-fit estimate of the 1.2\,mm continuum from 3C48A, is
probably also dominated by non-thermal emission. The 18\,cm map by
Feng et al.\ (2005) definitely shows non-thermal emission at the
position of 3C48A and the radio fluxes together with the 1.2~mm
estimate obey a power law that is consistent with synchrotron
emission. However, the finding of Zuther et al.\ (2004) that 3C48A is
highly reddened by dust suggests that 3C48A is also a source of dust
emission. This might be still negligible at 1.2~mm but becomes
important at submm wavelengths.

The 3.5\,mm continuum probably also has contributions from the VLBI
hot spot near the quasar, 3C48A, and the jet.  Although our 3.5\,mm
beam is too large to discriminate among these components, it seems
quite likely that at least the hot spot $B$ near the base of the jet
(Wilkinson et al.\ 1991 and Fig.~\ref{cont-ir}) and the extended jet
itself contribute to the 3.5\,mm flux. To estimate these
contributions, we adopted the 1.2\,mm model for the 3.5\,mm data, for
a spectral index of $-1$ between 3.5 and 1.2\,mm and smoothed it to
the beam at 3.5~mm. Figure\,\ref{model3.5mm} suggests that our 1.2\,mm
model also holds for the 3.5\,mm data.  Obviously, higher-resolution
observations are needed at 3.5\,mm and 1.2\,mm to confirm the three
continuum components we propose for 3C48.

   \begin{table}
     \centering
\caption[]{CO(1--0) positions, peak fluxes, and linewidths.}
     \begin{tabular}{ccccc}
          \hline
          \hline
           Epoch & RA & Dec. &CO(1--0) &Linewidth\\
	         & offset$^{\mathrm{b}}$ & offset$^{\mathrm{b}}$  
                 & peak flux
	   & FWHM \\
	         &  ($''$) & ($''$) & (mJy/beam) & (\kms)\\
         \hline
         1992$^{\mathrm{a}}$ &--- &--- & $\sim$7 & $\sim$250\\
         W97 &--- &--- & 9$\pm$2 & 270$\pm$20\\
         1995$^{\rm c}$ & \hspace{-0.4cm} $-$0.5$\pm0.6$ & 0.0$\pm0.6$ 
         & 9$\pm$2 & 240$\pm$40\\
         2003       & 0.5$\pm0.4$ & 0.0$\pm0.4$   & 6$\pm$2 & 330$\pm$50\\ 
         2003+1995 & 0.3$\pm0.3$ & 0.0$\pm0.3$ & 6$\pm$1   & 320$\pm$30 \\
%
            \hline
	 \end{tabular}
\begin{list}{}{}
\item[$^{\mathrm{a}}$] From Scoville et al.\ (1993).\\
\item[$^{\mathrm{b}}$] Relative to 01$^{\rm h}$37$^{\rm m}$41.30$^{\rm s}$,
	+33\degree09$'$35$''$ (J2000).\\
\item[$^{\mathrm{c}}$] From our re-reduction of the W97 data.\\
\end{list}
         \label{lineval}
   \end{table}

\subsection{Line emission}

\subsection{$^{12}$CO(1--0)}
\label{co10}
To allow for possible source variability, the continuum was subtracted
in the uv-plane separately for the 1995 and 2003 data sets. The line
peak fluxes in the W97 data and our new data agree within the errors
and with the value found by Scoville et al.\ (1993). The measured
linewidths and positions also agree within the errors among the
different data sets (see Table~\ref{lineval}). Thus, the final
(continuum-free) channel maps of the 1995 and 2003 data were finally
merged.

\begin{figure}[!]
   \centering 
   \resizebox{\hsize}{!}{\rotatebox{-90}{\includegraphics{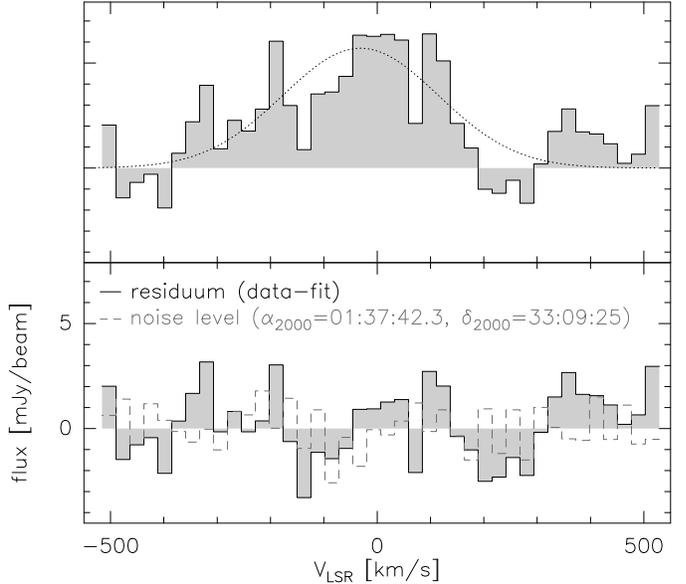}}}
   \caption{{\it Upper panel:} Spectrum at the CO(1--0) centroid in
3C48 taken from the merged 1995 and 2003 data set (solid histogram),
with a Gaussian fit to the data.  {\it Lower panel:} Residual spectrum
after subtracting the fit from the data (dotted line).  The dashed
histogram is a spectrum taken at an off-source position with no CO
emission, where the rms noise is 0.9\,mJy.  The velocity resolution is
26\,\kms.}
         \label{co-spec}
\end{figure}

Figure~\ref{co-chan} shows the merged maps of the CO(1--0) line
in 11\,MHz channels.  CO(1--0) emission is detected with a SNR of 4 to
5 in individual channels from $-$200\,\kms\ to +120\,\kms. A
single Gaussian fit to the spectrum (Fig.\,\ref{co-spec}) at the
centroid of the CO(1--0) emission yields the line parameters given in
Table~\ref{lineval}.  We then subtracted the fit profile, and found
that the residual spectrum had an rms noise of 1.9\,mJy, twice as high
as the measured rms noise of $\sim0.9$\,mJy at off-source positions.
The single Gaussian fit is thus probably too simple to account for the
observed line shape, but a better signal-to-noise ratio would be
needed for fitting multiple line components.

\begin{figure}[!]
   \centering 
   \resizebox{\hsize}{!}{\rotatebox{-90}{\includegraphics{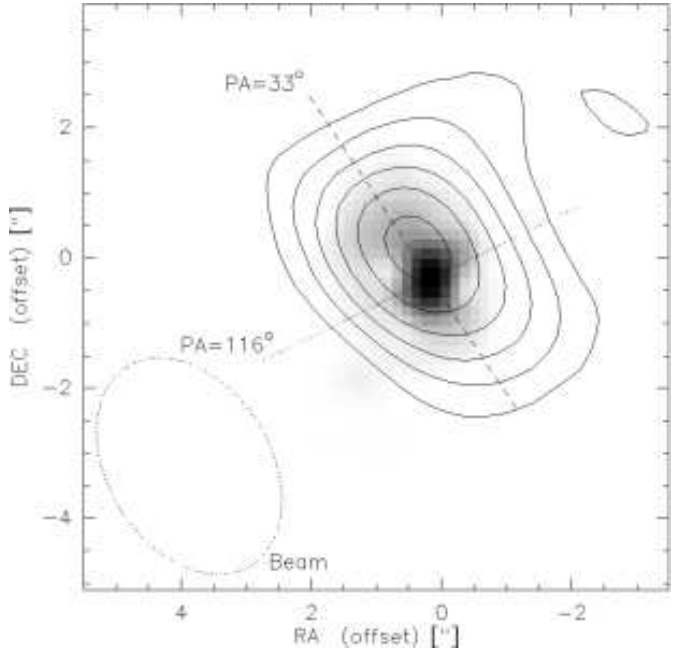}}}
      \caption{Integrated CO(1--0) emission of 3C48 (from
      $-220$\,\kms\, to 160\,\kms; merged 1995 and 2003 data). The
      synthesized beam is indicated in the lower left ({\it dotted
      ellipse}). The {\it dashed} and {\it dotted} lines shows the
      cuts along which the position-velocity diagrams were
      taken. Contour levels are from (3$\sigma$=)0.6 to
      1.6~Jy\,beam$^{-1}$\,\kms by 0.2~Jy\,beam$^{-1}$\,\kms. Beam:
      $3.6''\times 2.5''$ at PA=32\degree .  The CO map is superposed
      on the NIR image of Zuther et al.\ 2004 ({\it grey scale}).  }
         \label{co-all}
\end{figure}

The integrated emission over the whole velocity range (from $-$220 to
160\,\kms) is shown in Fig.~\ref{co-all}. The centroid of the total
CO(1--0) emission is located between the QSO and 3C48A. The emission
nicely covers the QSO and 3C48A and we find two extensions, one to the
north and one to the southwest, neither of which are aligned with the
major axis of the beam.  The position-velocity diagrams along the two
cuts\footnote{The cut at PA=33\degree (with respect to QSO-red) was
choosen to run along the velocity gradient seen in Fig.~\ref{ir-co}
and to hit still 3C48A and QSO-red (Fig.~\ref{co-comp}), while the cut
at PA=116\degree should go through QSO-red and QSO-blue
(Fig.~\ref{co-comp}). See next subsections for more details.}  shown
in Figs.~\ref{co-all} and \ref{co-comp} suggest three different
velocity components. A velocity gradient is visible over velocities
$-$120 to +80\,\kms. Two further components at $-$200\,\kms and
$+$120\,\kms appear to be unrelated to the ``inner'' velocity
gradient. Motivated by these diagrams, we made three different
integrated intensity maps over the respective velocity ranges
(Fig.\,\ref{co-comp}).  The positions of the CO components relative to
the infrared components led us to label the main CO feature, at
velocities $-$120\,\kms to $+$80\,\kms , as ``3C48A-CO'', and the
other two features as ``QSO-red'' ($+$120\,\kms) and ``QSO-blue''
($-$200\,\kms).  The 3C48A-CO component is centered close to the
3C48A-IR source, while the QSO-red-CO and QSO-blue-CO features are
roughly centered on the QSO-IR source, close to the QSO VLBI position.
All three CO components appear slightly extended in the interferometer
beam, but the signal-to-noise is not high enough to derive source
diameters.

\begin{figure}[!]
   \centering
   \resizebox{\hsize}{!}{\rotatebox{-90}{\includegraphics{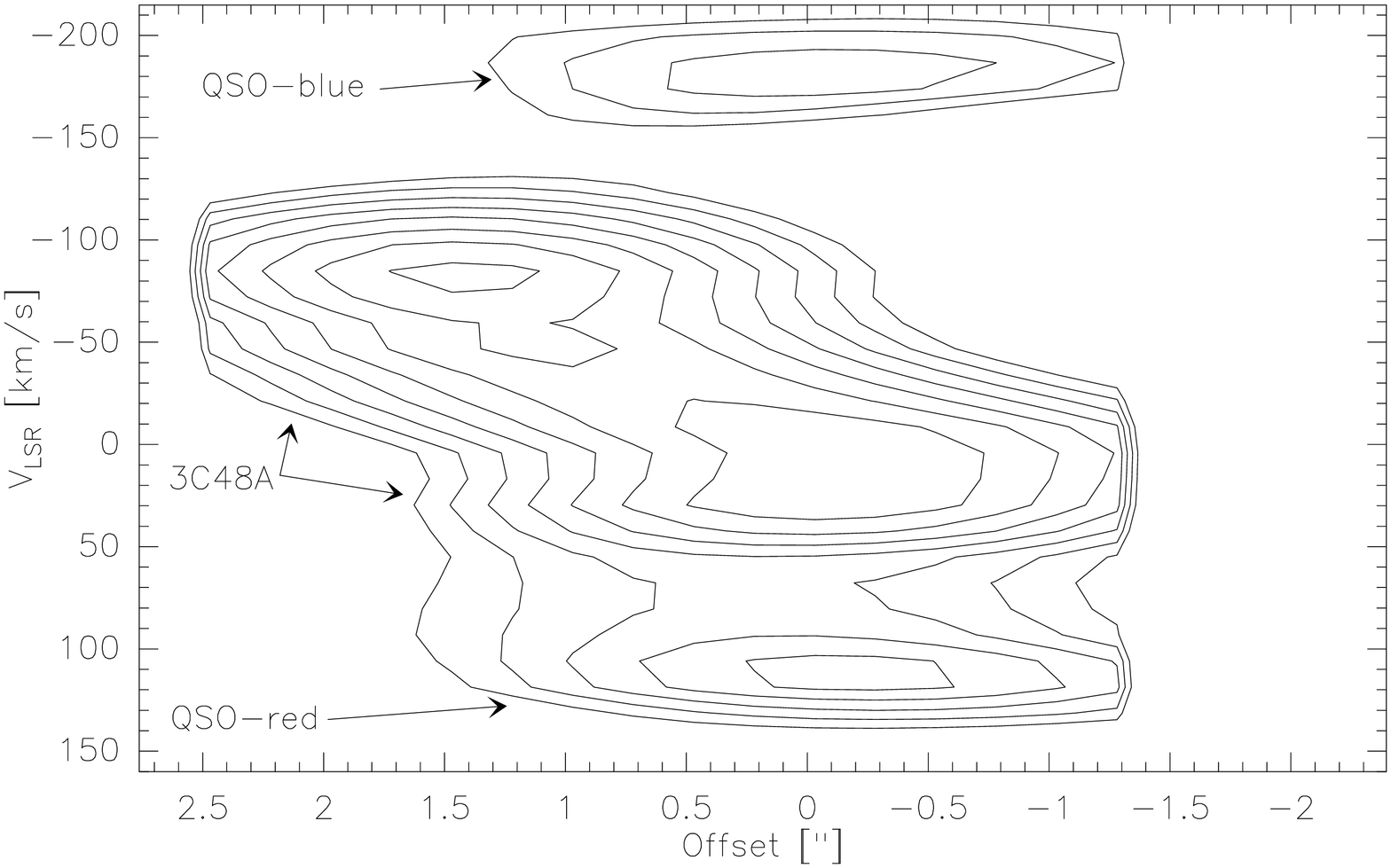}}}
   \resizebox{\hsize}{!}{\rotatebox{-90}{\includegraphics{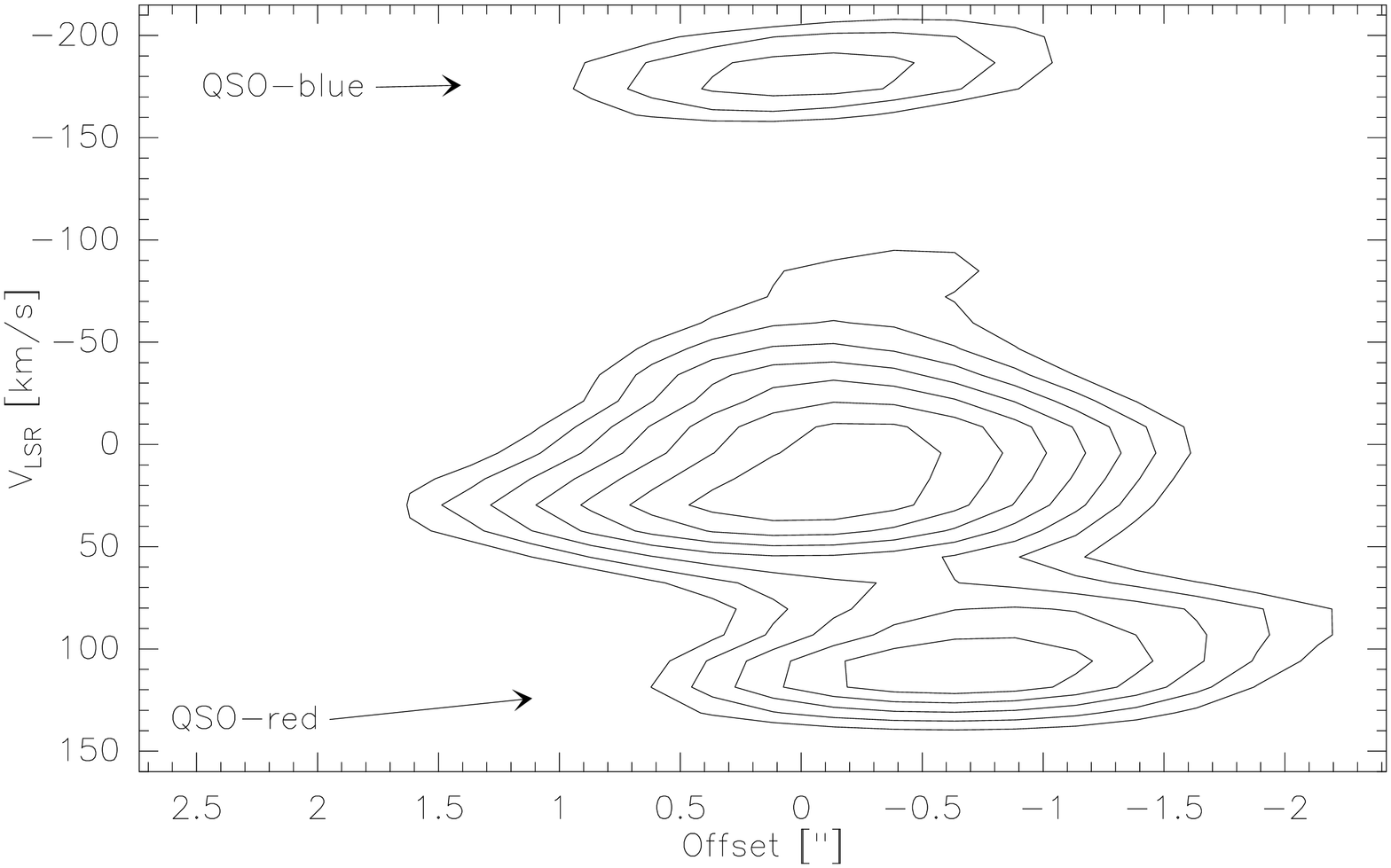}}}
      \caption{CO(1--0) position-velocity diagram along the cuts
      indicated by the dashed lines in Figs.~\ref{co-all} and
      \ref{co-comp} (merged 1995 and 2003 data set). Contours are
      from 35\% in steps of 5\% from the peak. {\it Upper panel:} The
      p-v cut from northeast to southwest, shown in Fig.~\ref{co-all}
      and in the middle panel of Fig.~\ref{co-comp}.  {\it Lower
      panel:} The p-v cut from southeast to northwest, shown in
      Fig.~\ref{co-all} and in the upper and lower panels in
      Fig.~\ref{co-comp}. The offsets are relative to the position of
      the CO-QSO-blue component.}
         \label{slice}
\end{figure}

\subsubsection{The 3C48A-CO velocity gradient}
At velocities corresponding to 3C48A-CO ($-$120\kms to +80\kms), the CO
centroid moves from south-west (positive velocities) to north-east
(negative velocities) in Fig.~\ref{co-chan}. Such a variation in
position was already suggested by W97 but their signal-to-noise ratio
was poor.  In our new 2003 data, this shift is highly
significant.  The velocity gradient is clearly visible in the
isovelocity map (Fig.~\ref{ir-co}, {\it right}) and in the pv-diagram
(Fig.~\ref{slice}).

\begin{figure}[!]
   \centering 
   \resizebox{7.5cm}{!}{\rotatebox{-90}{\includegraphics{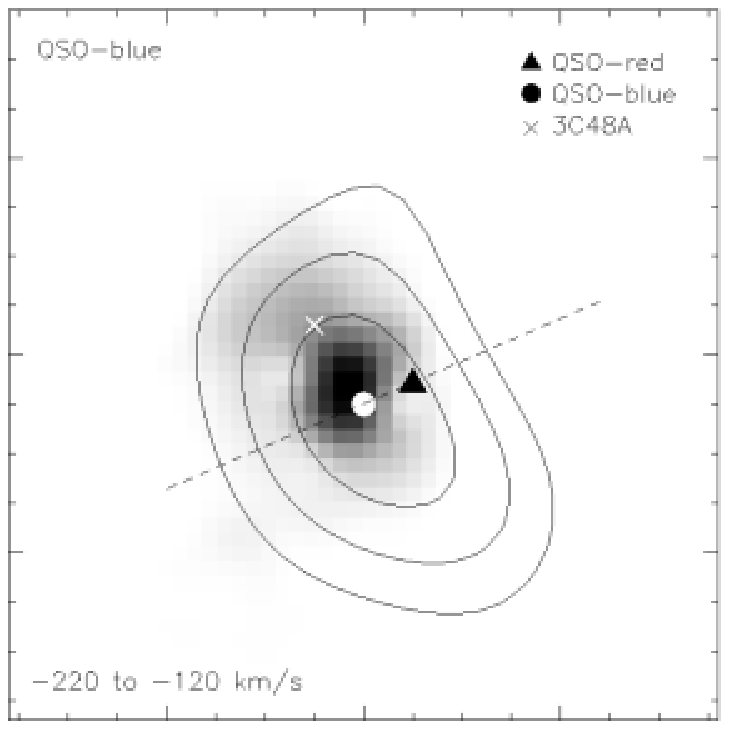}}}
   \vskip -0.4cm
   \resizebox{7.5cm}{!}{\rotatebox{-90}{\includegraphics{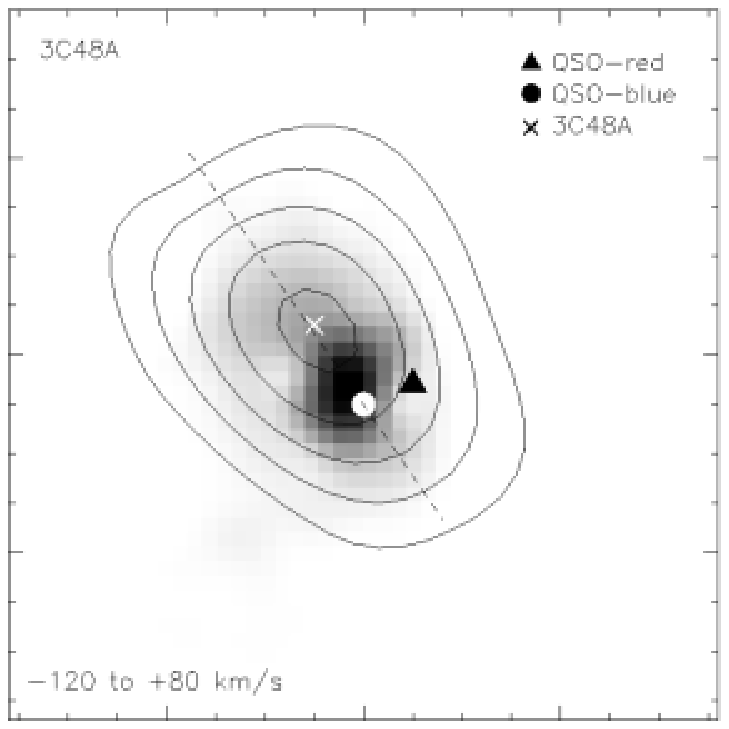}}}
   \vskip -0.4cm
   \resizebox{7.5cm}{!}{\rotatebox{-90}{\includegraphics{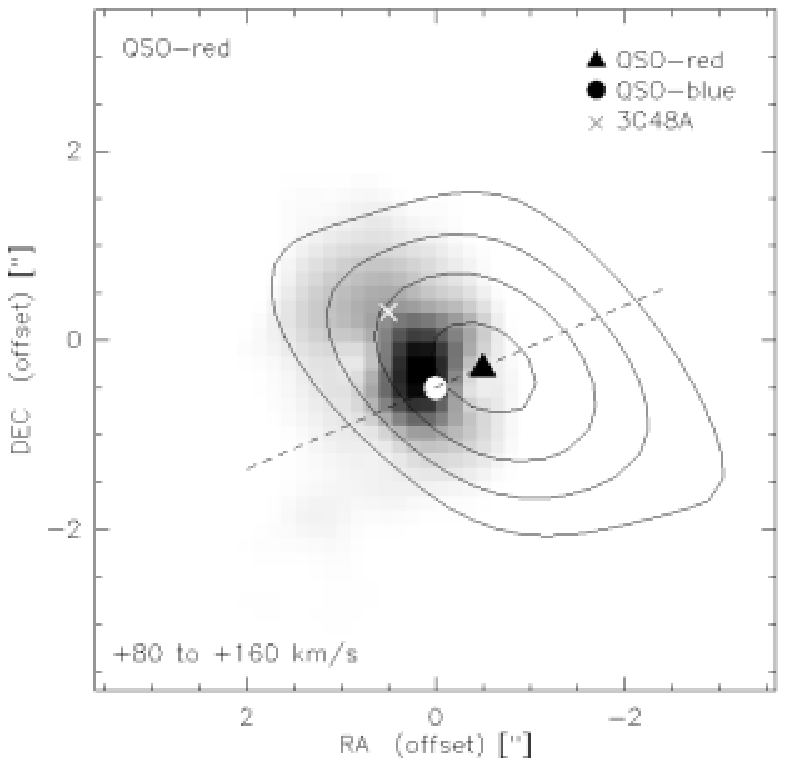}}}
      \caption{CO(1--0) integrated over the following ranges (merged
      1995 and 2003 data set): {\it Upper:} QSO-blue, from $-$220 to
      $-$120\,\kms ; contours run from (2$\sigma$=)0.08 to
      0.2\,Jy\,beam$^{-1}$\,\kms\ in steps of 1$\sigma$.  {\it
      Middle:} 3C48A, from $-$120 to 80\,\kms ; contours run from
      (3$\sigma=$)0.42 to 0.98\,Jy\,beam$^{-1}$\,\kms\ in steps of
      1$\sigma$.  {\it Lower:} QSO-red, from 80 to 160\,\kms ;
      contours run from (3$\sigma$=)0.15 to
      0.3\,Jy\,beam$^{-1}$\,\kms\ in steps of 1$\sigma$.  Dashed lines
      show the cuts on which the pv-diagrams were made.  The CO beam
      is $3.6''\times 2.5''$.  All the CO maps are superposed on the
      NIR image of Zuther et al.\ 2004 ({\it grey scale}).}
\label{co-comp} 
\end{figure}


\subsubsection{QSO-blue and QSO-red CO components}
In Figs.~\ref{co-chan}, \ref{slice}, and \ref{co-comp}, another two
compact CO features are seen at $-$180\,\kms and at +120\,\kms, that
we label as QSO-blue (at 5$\sigma$) and QSO-red (at 6$\sigma$). In
both the integrated intensity map (Fig.~\ref{co-comp}), and in the p-v
diagram (Fig.~\ref{slice}) along the cut in the upper and lower panels
of Fig.~\ref{co-comp}, the centroids of these two components differ by
$0.5''$.  Within the $0.3''$ positional uncertainties due to the low
signal-to-noise ratio, however, the two features roughly coincide with
each other and with the QSO.  We think these two components may arise
in a circumnuclear disk of molecular gas around the QSO.

   \begin{table}
     \centering
\caption[]{CO(1--0) luminosities and gas masses$^{\mathrm{a}}$.}
     \begin{tabular}{cccc}
          \hline
          \hline
CO        & Integrated    & Luminosity           & Gas mass \\
source    & CO flux       & $L^\prime_{\rm CO}$  & $M$(H$_2$+He)\\
component & (Jy\,\kms)    & (K\,\kms\,pc$^2$)   &(10$^{10}$\,\msun)\\
\hline
3C48A     & $1.2\pm 0.2$  &  8$\cdot$10$^9$      & 0.6 (3.6)$^\mathrm{b}$ \\
QSO-blue  & $0.3\pm 0.1$  &  2$\cdot$10$^9$      & 0.2 (0.9)$^\mathrm{b}$ \\
QSO-red   & $0.3\pm 0.1$  &  2$\cdot$10$^9$      & 0.2 (0.9)$^\mathrm{b}$ \\
\hline
Total     & $1.9\pm 0.2$  & 12$\cdot$10$^9$      & 1.0 (5.6)$^\mathrm{b}$ \\
\hline
	 \end{tabular}
\begin{list}{}{}
\item[$^{\mathrm{a}}$]{CO luminosities and gas masses are for 
$H_0=71$\,\kms\,Mpc$^{-1}$, $\Omega_m = 0.27$,
 and $\Omega_\Lambda = 0.73$, which yield an angular diameter distance of 
3C48 of 1.044\,Gpc. 1$''$ corresponds to 5.06\,kpc.}\\
\item[$^\mathrm{b}$]{The gas masses were derived using
  0.8\,(K\,\kms\,pc$^2$)$^{-1}$ (see text), while the values in
  brackets are estimated via the standard conversion factor of
  4.6\,(K\,\kms\,pc$^2$)$^{-1}$.}\\
\end{list}
         \label{masses}
   \end{table}

\section{Gas mass and dynamical mass in 3C48}
Table~3 lists the gas masses (H$_2$ plus helium) that we estimate from
the CO(1--0) luminosities, using the mean conversion factor of
0.8\,\msun\,(K \kms\,pc$^2)^{-1}$ obtained by Downes \& Solomon (1998)
by kinematic/radiative-transfer modeling of the molecular gas in the
circumnuclear regions of Ultra-Luminous InfraRed Galaxies (ULIRGs).
This value is a factor of six lower than the value that would hold in
self-gravitating molecular clouds in spiral arms of the Milky Way, and
is relatively insensitive to assumed [CO]/[H$_2$] abundances, because
the CO is opaque.  As a check, one may estimate a {\it lower} limit on
the molecular gas mass, by assuming the CO is optically thin (Solomon
et al.\ 1997). This method yields a lower limit to the (total) 3C48
gas mass of $> 8\times10^9$\,\msun .  An independent estimate of the
gas mass can be made from the optically thin dust flux.  For a dust
flux density of 0.55\,Jy at $\nu_{\rm obs}$=1.62\,THz (Meisenheimer et
al.\ 2001) and a rest-frame dust temperature $T_d$=50\,K, as in
Fig.\ref{sed}, we obtain a dust mass of $M_{\rm dust}\simeq 1.6\times
10^8$\,\msun.  For a gas-to-dust mass ratio of 150, the gas mass would
be $M_{\rm gas}\sim 2\times10^{10}$\,\msun , of the same order as the
molecular gas mass estimate from the CO luminosity.

Without size measurements, one cannot obtain reliable values for the
dynamical mass (gas plus stars) within the CO-emitting regions.
Representative estimates of $R V^2/G$ would be
3.2$\times10^{10}$\,\msun\ for an 0.5$''$-diameter (1.3~kpc radius)
circumnuclear disk rotating at 330\,\kms\ around the 3C48 quasar, or
6.4$\times$10$^{10}$\,\msun\ for the 3C48A and 3C48 nuclei, if their
true separation is 5\,kpc ($1''$), and they are orbiting their center
of mass at a radius of 2.5\,kpc at a velocity of 330\,\kms\ (the CO
linewidth in Table~2).  The latter dynamical mass estimate is close to
the estimate of the gas mass, but could easily be much higher if the
true distance is greater than the projected separation of 3C48 and
3C48A on the sky, and/or if the relative velocity of the merger nuclei
is greater than $330$\,\kms. Thus, the estimate of the dynamical mass
should be taken as lower limit.

\begin{figure*}[!]
   \centering 
\resizebox{9cm}{!}{\rotatebox{-90}{\includegraphics{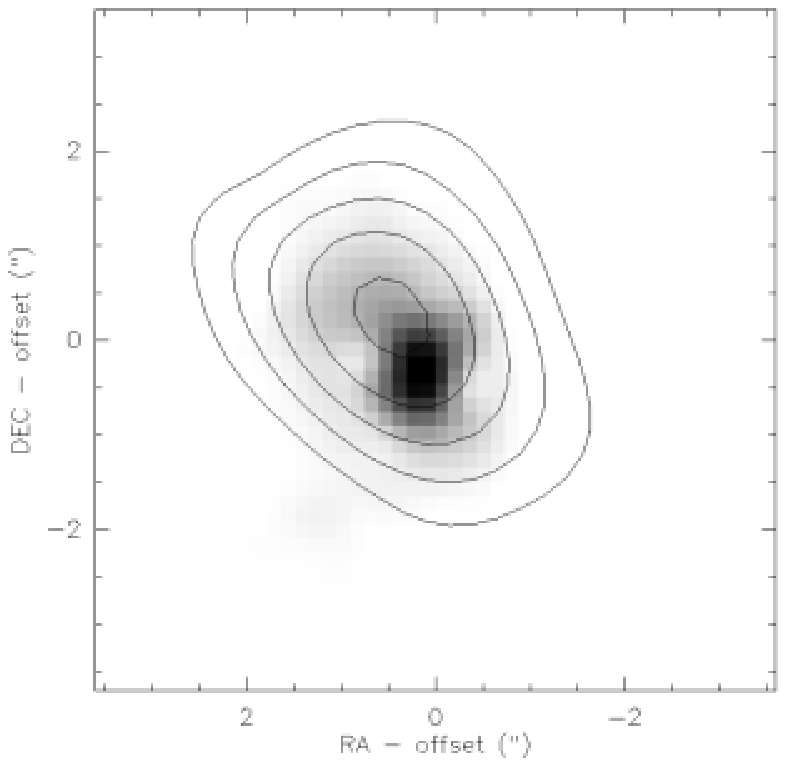}}}
\resizebox{8cm}{!}{\rotatebox{-90}{\includegraphics{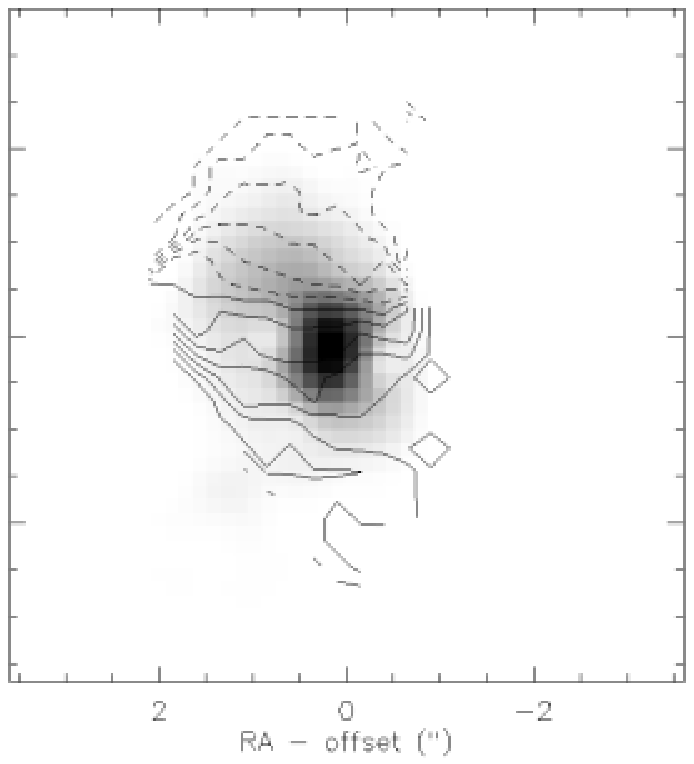}}}
      \caption{ {\it Left:} 3C48 Integrated CO(1-0) emission from
$-$120 to +80\,\kms . Contours are 0.2 to 0.8\,Jy\,beam$^{-1}$\kms by
0.1\,Jy\,beam$^{-1}$\kms .  The CO beam is $3.5''\times 2.6''$.  {\it
Right:} CO isovelocity map.  Velocity contours run from $-$95 to
$-$45\,\kms\ ({\it dashed}) and $-$35 to 25\,\kms\ ({\it solid}) in
steps of 10\,\kms .  Both CO maps are superposed on the NIR image of
Zuther et al.\ 2004 ({\it grey scale}). Data are taken from the merged
1995 and 2003 data sets}.
         \label{ir-co}
\end{figure*}

\section{3C48 and other sources  where CO has been observed in 
two merger nuclei} In the scenario of Sanders et al.\ (1988), galaxy
interactions and mergers trigger the formation of ULIRGs, which evolve
to turn on the activity of the massive black holes at the centers of
quasars and radio galaxies.  With our new CO evidence for two
circumnuclear molecular disks, 3C48 now joins a growing list of
powerful AGNs where CO is detected in the merger partners.  At low
redshifts these include the sample of quasars and ULIRGs observed in
CO by Evans et al.\ (2001, 2002), some of which have double nuclei.
At high redshifts, prominent examples are the two CO systems detected
in the powerful radio galaxy 4C\,41.17 (De~Breuck et al. 2005), the
two CO systems in the radio galaxy 4C\,60.07 (Papadopoulos et al.,
2000; Greve et al. 2004), the double CO sources in the quasars BRI
1202-0725 and BRI 1335-0417 (e.g., Carilli et al.\ 2002), the two CO
nuclei in the $z=6.4$ quasar J1148+52 (Walter et al. 2004) and the two
optical/IR objects L1 and L2 in the quasar SMMJ02399+0256 (Ivison et
al.\ 1998). For this last quasar, we think the two CO velocity peaks,
first detected by Frayer et al.\ (1998), correspond to the two
circumnuclear disks of the merger, rather than a single large disk as
proposed by Genzel et al.\ (2003).

Because most luminous, low-redshift QSOs appear to be in gas-rich host
galaxies (Scoville et al. 2003), it is worth reviewing the two
best-known nearby mergers of gas-rich galaxies, and their effects on
the molecular gas --- Arp\,220 at a distance of 75\,Mpc, and the
``Antennae'' galaxies (NGC~4038/39) at a distance of 18\,Mpc.  Because
they have been so well-studied, the tidal tails of the Antennae
galaxies were used by Scharw\"achter et al.\ (2004) to simulate
possible tidal tails, viewed from a different angle, in 3C48.  The
optical morphology (e.g. Whitmore et al.\ 1999), shows the Antennae
galaxies to be an early-stage merger of two gas-rich spiral galaxies.
Gao et al.\ (2001) derive a molecular gas mass of
($\sim1.5\times10^{10}$\msun), extended over both galaxies.  There is
a large amount of molecular gas of $\sim4\times10^{9}$\msun\ in the
overlapping region of the two galaxies (Zhu et al.\ 2003).  The CO,
the mid- and far-infrared, and the submm- and cm-radio continuum all
peak in the region between the two merging disks. The large IR
luminosity of $\sim 10^{11}$\solar\ puts the Antennae galaxies in the
LIRG class (Sanders \& Mirabel 1996). From long-wavelength studies
(e.g. Hummel \& van der Hulst 1986; Mirabel et al.\ 1998; Haas et al.\
2000; Neff \& Ulvestad 2000), it is clear the IR luminosity of the
Antennae galaxies is due to system-wide star formation, not an
AGN. The two nuclei $\sim$7kpc projected separation of the two nuclei
is about the same as that of 3C48 and 3C48A.

Arp\,220 has a high molecular gas content of $\sim 10^{10}$\msun\
within the central kiloparsec (Scoville et al.\ 1986). It also
contains two nuclear components with a projected separation of
300\,pc, and an extended tidal tail that led to the hypothesis of an
ongoing merger (e.g., Norris 1985; Graham et al.\ 1990). The huge IR
luminosity ($L_{\rm 8-1000\mu m}=1.4 \times10^{12}$\solar) puts it in
the ULIRG class (Soifer et al.\ 1987).  Scoville et al.\ (1998) report
on a high near-IR obscuration of one of the two nuclei, as is the case
in 3C48A (Zuther et al. 2004). Besides a large molecular gas disk
($r\sim 1$kpc) rotating around the dynamical center of the system,
high resolution ($\sim 0.5''$) observations of the CO emission unveil
nuclear disks ($r\sim$100pc) around both nuclei (Downes \& Solomon
1998). These two nuclear disks appear to rotate orthogonally with
respect to each other and have molecular gas masses of 10$^9$\msun\, and
dynamical masses of $\gtrsim2\times10^9$\msun\ (Sakamoto et al.\
1999).  Eckart \& Downes (2001) showed that the Arp\,220 CO kinematics
may also be interpreted as a single, warped disk.  Except for the
scale, the overall situation in 3C48 resembles that in Arp\,220.  Our
CO data in this paper suggest that there are also two rotating,
molecular gas disks in 3C48.  The main difference is that 3C48 is a
powerful quasar, while no obvious AGN has been identified in Arp\,220.

\section{Summary and conclusions}
1) Our new CO(1--0) results show that the main part of the emission
--- the central part of the CO line actually comes from 3C48A, not the
quasar.  The higher sensitivity of the new CO data shows a clear
velocity gradient across 3C48A, indicating rotation of a disk of
molecular gas at 3C48A, with an extension toward the north and
southwest.  This CO concentration is the strongest argument supporting
the idea that 3C48A is a second nucleus.

2) The data clearly indicate two different dynamical systems in the
molecular gas: the extended disk toward 3C48A, and a second,
independent gas reservoir to the southwest, around the QSO itself.

3) The total molecular gas mass of a few times $10^{10}$\,\msun\, is
typical of the circumnuclear disks in advanced-merger ULIRGs.

4) The 1.2\,mm nonthermal continuum was mapped for the first time at
resolution of $\sim0.8''$, and observed to be extended.  At 1.2\,mm,
the continuum is clearly elongated towards the second NIR nuclear
component 3C48A and the radio jet.  These extensions are consistent
with that on the 18\,cm MERLIN map (Akujor et al.\ 1994).  The 3.5\,mm
continuum data also suggest extended emission towards 3C48A.

5) Model fits to the 1.2\,mm continuum suggests three distinct
components: one at the 3C48 QSO to the southwest, one along the
extended jet to the north and a third one at the position of the
candidate second merger nucleus 3C48A to the north-east.  While the
1.2\,mm components corresponding to both the QSO hot spot and the
radio jet are synchrotron emission, the 3C48A component may have a
partial contribution from thermal dust emission at 1.2\,mm.
Higher-resolution mm and submm continuum observations (e.g.\ with the
SMA or the upgraded PdBI) are needed to better separate the components
and to analyse the contribution of thermal and non-thermal emission.

6) In the 1.2\,mm continuum, 3C48A is located to the east of the jet,
  i.e.\ does not align with the jet.  This shows that the jet does not
  hit 3C48A head-on.  A possible but still speculative explanation
  might be that the non-collimated and diffuse jet has been disrupted
  a first time by the dense circumnuclear disk around the quasar
  itself, and then diverted a second time, to the north, by the
  magnetic field pressure associated with 3C48A (see the 1.66\,GHz map
  by Wilkinson et al. 1991, their Fig.~1).

7) 3C48 joins a growing list of quasars in which \emph{two} merger
nuclei have been detected in CO. 3C48 shows similar properties to
these double-CO-nuclei quasars at low and high redshift, and also to
the molecular gas in nearby non-quasar mergers like Arp\,220 and the
Antennae galaxies.  All these objects have high molecular gas masses
of a few times $10^{10}\,$\msun, high infrared luminosities and have
two components with different projected separations interpreted as two
nuclei.

\begin{acknowledgements}
      We thank T.L.\, Wilson \& S.\,Guilloteau for kindly 
      providing their earlier data (Wink, Guilloteau \&
      Wilson 1997, W97) and for helpful discussion. Part of this work
      was supported by the German
      \emph{Son\-der\-for\-schungs\-be\-reich, SFB\/,} project number
      494. This paper was based on observations with the IRAM
    Interferometer. IRAM is supported by INSU/CNRS
   (France), MPG (Germany) and IGN (Spain).
\end{acknowledgements}

\end{document}